# Methane Oxidation to Methanol without $CO_2$ Emission: Catalysis by Atomic Negative Ions

Aron Tesfamichael[1], Kelvin Suggs[1], Zineb Felfli[2], and Alfred Z. Msezane[2]
[1]Department of Chemistry, Clark Atlanta University, Atlanta, Georgia 30314, USA
[2]Department of Physics and Center for Theoretical Studies of Physical Systems, Clark Atlanta University, Atlanta, Georgia 30314, USA

**ABSTRACT**
The catalytic activities of the atomic $Y^-$, $Ru^-$, $At^-$, $In^-$, $Pd^-$, $Ag^-$, $Pt^-$, and $Os^-$ ions have been investigated theoretically using the atomic $Au^-$ ion as the benchmark for the selective partial oxidation of methane to methanol without $CO_2$ emission. Dispersion-corrected density-functional theory has been used for the investigation. From the energy barrier calculations and the thermodynamics of the reactions, we conclude that the catalytic effect of the atomic $Ag^-$, $At^-$, $Ru^-$, and $Os^-$ ions is higher than that of the atomic $Au^-$ ion catalysis of $CH_4$ conversion to methanol. By controlling the temperature around 290K ($Os^-$), 300K ($Ag^-$), 310K ($At^-$), 320K ($Ru^-$) and 325K ($Au^-$) methane can be completely oxidized to methanol without the emission of $CO_2$. We conclude by recommending the investigation of the catalytic activities of combinations of the above negative ions for significant enhancement of the selective partial oxidation of methane to methanol.

1. **Introduction**

The conversion of methane into valuable products is of considerable industrial, economic and environmental interest as well as of great scientific importance. However, a great challenge to overcome is that in the absence of an appropriate catalyst methane undergoes complete combustion, yielding carbon dioxide and water at around 340 K with very little competition for the formation of useful products that can occur at elevated temperatures. Recently, theoretical evidence has been established demonstrating the possibility of producing methanol from methane without $CO_2$ emission through the use of the atomic $Au^-$ ion catalyst [1]. When the atomic $Au^-$ ion catalyst is employed methane undergoes oxidation to methanol at 325 K, a temperature that is below the generation of $CO_2$. The effect of the atomic $Au^-$ ion is to lower the transition state by about 32% compared to the case of the absence of the catalyst for the complete oxidation of methane to methanol without the $CO_2$ emission [1].

The role of atomic particles and nanoparticles in catalysis continues to attract extensive investigations from both fundamental and industrial perspectives [2-16]. Recently, we have added the novel atomic negative ions to the study of catalysis at the atomic scale by performing transition state calculations using dispersion-corrected density-functional theory for the following reactions: 1) Conversion of $H_2O$, HDO, and $D_2O$ to $H_2O_2$, $HDO_2$, and $D_2O_2$, respectively using atomic $Au^-$ and atomic $Pd^-$ ion catalysis [17, 18] and 2) Complete and partial oxidation of methane in the absence and presence of the atomic $Au^-$ ion [1].

Our results indicated that the atomic Au⁻ ion catalyzes excellently; however, the atomic Pd⁻ ion has a higher catalytic effect on the formation of peroxide from light, intermediate, and heavy water [17, 18], by 35%, 50% and 62%, respectively, consistent with the recent experimental observations [8, 16]. The fundamental mechanism of negative ion catalysis in the oxidation of water to peroxide catalyzed by the atomic Au⁻ ion has been attributed to the anionic molecular complex Au⁻$(H_2O)_{1, 2}$ formation in the transition state, with the atomic Au⁻ ion breaking up the hydrogen bond strength in the water molecules, permitting the formation of the peroxide in the presence of $O_2$ usually provided by the support. Similarly, in the conversion of methane to methanol using the atomic Au⁻ ion, the anionic molecular complex Au⁻$(CH_4)$ formation weakens the C-H bond in the transition state.

The experiment [11] determined the vertical detachment energies (VDEs) of the Au⁻M complexes (M = Ne, Ar, Kr, Xe, $O_2$, $CH_4$ and $H_2O$). Importantly it also found a stronger interaction between the atomic Au⁻ ion and $H_2O$ and that the atomic Au⁻ ion reacts well with $CH_4$ but weakly with $O_2$ and the noble gases. Furthermore, the anionic complex Au⁻$(H_2O)_2$ has been characterized as atomic Au⁻ interacting with two water molecules, *i.e.* as the anion-molecule complex [19]; a similar analysis applies to the anionic complex Au⁻$(CH_4)$. The large electron affinity of the Au atom played the essential role; it is important in the dissociation of the Au⁻$(H_2O)_2$ anionic complex breaking into the Au anion and $H_2O$ [19]. Essentially, these two experiments [11, 19] probed nanogold catalysis at its most fundamental level. Their results are important in the fundamental understanding of the significant catalytic properties of gold and palladium nanoparticles.

The fundamental atomic physics mechanism responsible for the oxidation of water to peroxide and of methane to methanol using the atomic Au⁻ ion catalysis has been attributed to the interplay between resonances and Ramsauer-Townsend (R-T) minima in the electron elastic total cross sections (TCSs) for the Au atom, along with its large electron affinity (EA) [20, 21]. Ramsauer-Townsend minima, shape resonances and dramatically sharp long-lived resonances generally characterize the near-threshold electron elastic scattering TCSs for both the ground and excited states of simple and complex atoms [22-24]. The long-lived resonances have been identified with the formation of stable bound states of the relevant negative ions formed during the elastic collision between the slow electron and the target neutral atom as Regge resonances [22]. The recently developed Regge-pole methodology wherein the crucial electron-electron correlations are embedded has been employed for the calculations [25]. The vital core polarization interaction is incorporated through the well-investigated Thomas-Fermi type potential [26, 27].

Reliable atomic and molecular affinities, manifesting the existence of long-lived negative ion-formation, are crucial to the understanding of a large number of chemical reactions involving negative ions [28]. The role of resonances is to promote anion formation through electron attachment [29]. Recently, using the Regge pole methodology [25], we investigated electron elastic scattering TCSs for Y, Ru, At, In, Pd, Ag and Pt atoms in the electron impact energy range $0 \leq E \leq 7.0$ eV in search of atomic negative ion catalysts [30]. The configuration of the resonances and the R–T minima in the atomic Au TCS was used as the benchmark for atomic negative ion catalysts. We concluded from the electron elastic TCSs for all the above atoms but In, that they represent excellent candidates for atomic negative ion catalysts individually or in various combinations. This investigation has been motivated further by the finding that the catalytic activity of the Au–Ag–Pd trimetallic nanoparticles is efficient in catalyzing the Heck reaction [13] and that the methanol oxidation current of the ternary Pt–Ru–Ni catalyst increased significantly in comparison with that of the binary Pt–Ru catalyst [14].

The purpose of this paper is to investigate which atomic negative ion catalysts from the above candidates perform better than the atomic $Au^-$ ion catalyst in the oxidation of methane to methanol without $CO_2$ emission into the atmosphere. Toward this end the following studies have been performed: 1) Electron elastic TCSs calculations using the Au TCS as the benchmark for an atomic negative ion catalyst; 2) Transition state calculations using dispersion-corrected density-functional theory; and 3) Thermodynamics properties analysis. We note that the electron elastic TCSs used here were taken from [30, 31, 32].

## 2. Reactions and Calculation Methods

### 2.1 Reactions

Here we investigate which catalysts among the atomic $Y^-$, $Ru^-$, $At^-$, $In^-$, $Pd^-$, $Ag^-$, $Os^-$, and $Pt^-$ ions perform better than the atomic $Au^-$ ion in the catalytic oxidation of methane to methanol without the $CO_2$ emission.

Following [20] in this study we first consider the slow oxidation of methane to methanol without the atomic negative ion catalyst, namely

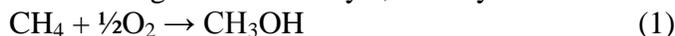
$$CH_4 + \tfrac{1}{2}O_2 \rightarrow CH_3OH \qquad (1)$$

Then we apply the atomic $Au^-$ ion to speed up the reaction (1) and obtain

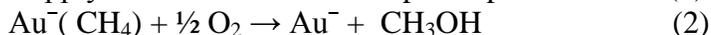
$$Au^-(CH_4) + \tfrac{1}{2}O_2 \rightarrow Au^- + CH_3OH \qquad (2)$$

and

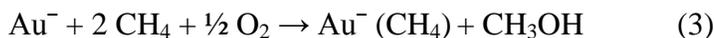
$$Au^- + 2\,CH_4 + \tfrac{1}{2}O_2 \rightarrow Au^-(CH_4) + CH_3OH \qquad (3)$$

Add reactions (2) and (3) and obtain

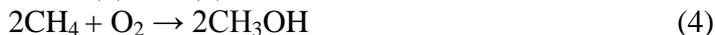
$$2CH_4 + O_2 \rightarrow 2CH_3OH \qquad (4)$$

We note that the atomic $Au^-$ ion is freed (not consumed) in the reaction (4), consistent with the role of a catalyst. Similar results as in (4) are obtained when the remaining negative ions are employed.

Figure 1 represents the catalytic cycle for the reactions (2) and (3) using the atomic $Au^-$ ion catalyst (similar results are obtained for the remaining negative ion catalysts). In reaction (3), the $CH_4$ molecule attaches to the atomic $Au^-$ ion catalyst forming the $Au^-(CH_4)$ anionic molecular complex, breaking the significant C-H bond, thereby allowing the formation of $CH_3OH$ in the presence of $O_2$ (the $O_2$ hardly reacts with the atomic $Au^-$ ion, see [11, 19]). This is the cycle on the right hand side of the Fig. 1, read from top to bottom. After the reaction, the $Au^-(CH_4)$ complex combines with the $O_2$ forming the $CH_3OH$, thereby freeing the atomic $Au^-$ ion catalyst for the next cycle (left hand side of Fig. 1, read from the bottom to top).

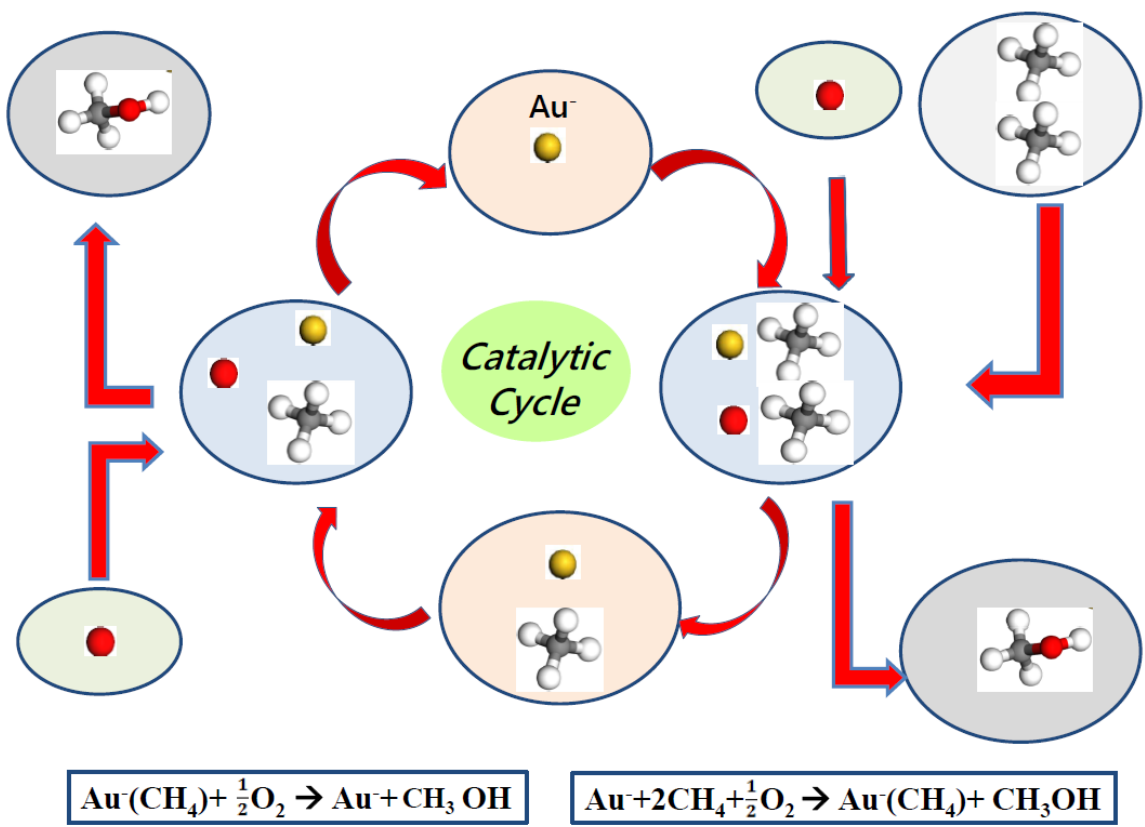

**Figure 1:** Catalytic cycle of the oxidation of methane to methanol without $CO_2$ emission through the atomic $Au^-$ ion catalyst. Similar catalytic cycles for the other relevant negative ions can be constructed.

## 2.2 Calculation Methods

### 2.2.1 Electron Elastic Total Cross Sections and Electron Affinity Calculations

In order to determine the appropriateness of an atom and its anion for use as nanocatalysts, we must first map and delineate the atom's resonance structure at low electron impact energy. Electron-electron correlations and core polarization interactions are essential for the stability of most atomic negative ions. So, we have used our Regge pole, also known as the complex angular momentum (CAM), method [25] to calculate accurate low energy electron scattering TCSs for the atoms Y, Ru, Pd, Ag, and Pt [30] and At, and In [31] in the electron impact energy range $0 \leq E \leq 7.0$ eV as well as for the Os atom [32] and extracted the attendant EAs. The choice of these atoms is dictated by the finding that the negative ions of the Ag, At, Ru and Os atoms catalyze the oxidation of methane to methanol without the $CO_2$ emission better than the atomic $Au^-$ ion (see results below). The TCSs for these atoms are characterized by R-T minima, shape resonances and dramatically sharp resonances, corresponding to the bound states of the negative ions formed during the collision. Most important here is that the EAs of these atoms are relatively large, around 2.0 eV and are in the deep second R-T minima.

Table 1 summarizes the essential data for the various atoms of interest here. We note that a combination of two or more of these atoms extends the effective second minimum significantly so that the catalytic properties of two or more negative ions can be enhanced compared to the case when only one is used. Furthermore, the second R-T minima for atomic Ag and At are deepest; their profound effect will be seen under results. From Table 1, we note the dramatic change in the EA from Z=39 through Z=44; this manifests the sensitivity of the EA values to the atomic structure configurations. From the TS values the percentage catalytic effects of the various atomic negative ions have been evaluated and the results are presented in the last column. From the Table 1 it is clear that the atomic $Au^-$ ion reduces the TS by 31.7%, while the $Ru^-$, $At^-$, $Ag^-$, and $Os^-$ ions reduce the TS by 32.7%, 34.7%, 36.7% and 39.0%, respectively.

**Table 1:** Calculated electron affinities (EAs), transition state (TS), Ramsauer-Townsend minima (R-T), all in eV and temperature T(K) for the various atoms of interest.

| Atomic Negative Ion | Z | 1st R-T Minimum | EA | 2nd R-T Minimum | TS | T(K) at $\Delta G=0$ | Catalytic Effect (%) |
|---|---|---|---|---|---|---|---|
| No Catalyst | | | | | 4.41 | 475 | 0.0 |
| $Au^-$ | 79 | 0.692 | 2.262 | 2.057 | 3.01 | 325 | 31.7 |
| $Y^-$ | 39 | 0.652 | 1.754 | 2.103 | 3.62 | 390 | 17.9 |
| $Ru^-$ | 44 | 0.849 | 2.265 | 2.897 | 2.97 | 320 | 32.7 |
| $At^-$ | 85 | 0.810 | 2.510 | 2.490 | 2.88 | 310 | 34.7 |
| $In^-$ | 43 | 0.066 | 0.380 | - | 3.95 | 425 | 10.4 |
| $Pd^-$ | 46 | 0.930 | 1.948 | 3.134 | 3.71 | 400 | 15.9 |
| $Ag^-$ | 47 | 0.970 | 2.240 | 3.347 | 2.79 | 300 | 36.7 |
| $Pt^-$ | 78 | 0.672 | 2.163 | 2.166 | 3.62 | 390 | 17.9 |
| $Os^-$ | 76 | 0.631 | 1.910 | 1.826 | 2.69 | 290 | 39.0 |

### 2.2.2 Transition State Calculation

We have employed the first principles calculations based on Density Functional Theory (DFT) and dispersion corrected DFT approaches [33] for the investigation of the transition states. For geometry optimization of the structural molecular confirmation we utilized the gradient-corrected Perdew-Burke-Ernzerfof (PBE) parameterizations [34] of the exchange-correlation rectified with the dispersion corrections [35]. The double numerical plus polarization basis set was employed as implemented in the DMol3 package [36]. The dispersion-correction method, coupled to suitable density functional, has been demonstrated to account for the long-range dispersion forces with remarkable accuracy. We used a tolerance of $1.0 \times 10^{-3}$ eV for energy convergence. A transition-state search employing nudged elastic bands facilitates the evaluation of energy barriers [37-39]. Finally, the energy of the transition state was calculated and the thermodynamic properties of the reaction were analyzed using the DMol3 package [36]. The results are displayed in Figures 2.

### 2.2.3 Thermodynamics Properties of Reactions

The mechanism for creating and breaking bonds can be understood also from a theoretical chemistry perspective. For example, in the $H_2O_2$ catalysis from water using the atomic $Au^-$ negative ion the H-bond breaking mechanism has been attributed to the formation of the anionic $Au^-(H_2O)_2$ complex, while in the oxidation of $CH_4$ to methanol, the C-H bond breaking has been attributed to the formation of the anionic complex $Au^-(CH_4)$. Bond-breaking has a direct effect on the change in the Gibbs free energy, G (ΔG = ΔH -TΔS) where H, T, and S represent enthalpy, temperature, and entropy, respectively.

When the atomic $Au^-$ ion catalyst is introduced into the oxidation of methane, the breaking of the C-H bonding in $CH_4$ results. Therefore, the system changes from relative order to less order. Hence, the entropy of the system increases, whereas the enthalpy of the system decreases. The overall result is a negative Gibbs free energy and the process results in the spontaneous formation of methanol. To gain a deeper understanding of the process of negative ion catalysis we have also calculated the rate of a reaction using Arrhenius equation [40] and compared the number of molecules that can react in the absence and presence of an anion catalyst at constant (room) temperature using the expression

$$K = A \exp(-E_a/RT), \qquad (5)$$

where $K$ is the rate constant, T is the temperature in Kelvin, R is the gas constant (8.31 J-mol/K), $E_a$ is the activation energy in J/mole and $A$ is the frequency factor which includes factors such as the frequency of collisions and their orientation. It varies slightly with temperature, although not much. It is often taken as constant across small temperature changes.

## 3. Results

Figures 2, present the optimized structures of the reactants, transition states (TSs), and products (EPs) of the oxidation of $CH_4$ leading to the formation of $CH_3OH$ using the atomic negative ion catalysts from Au, Y, Ru, At, In, Pd, Ag, Pt, and Os. The data in figure 2(A) correspond to the absence of a catalyst while those in 2(B), 2(C), 2(D), 2(E), 2(F), 2(G), 2(H), 2(I), and 2(J) are data when the $Au^-$, $Y^-$, $Ru^-$, $At^-$, $In^-$, $Pd^-$, $Ag^-$, $Pt^-$, and $Os^-$ ion catalysts are present, respectively. The red, white and grey spheres represent the O, H and C atoms, respectively; the gold, green, black, blue, pink, light green, purple, dark purple, and brown spheres represent the $Au^-$, $Y^-$, $Ru^-$, $At^-$, $In^-$, $Pd^-$, $Ag^-$, $Pt^-$, and $Os^-$ ions, respectively.

The TS and EP, both in eV with the errors of calculation ± 0.01eV, represent respectively the calculated transition state energy and the energy of the products. Recently, our group performed the catalytic oxidation of methane to methanol without the emission of the $CO_2$. Using the $Au^-$ ion as the catalyst essentially disrupts the C-H bonding in $CH_4$ oxidation thereby eliminating the competition from the carbon dioxide formation. We have concluded by recommending that the negative ions of the atoms such as those in [30, 31] be investigated for possible catalytic activities in the selective partial oxidation of methane. The proposed fundamental mechanism involves the breaking of the stable C-H bonds in the methane molecule in the transition state through the formation of the anionic $X^-(CH_4)$ complex, where X represents one of the above atomic catalysts. The role of the $X^-$ ion is to disrupt the stable C-H bonds in the methane molecule, allowing the formation of the methanol in the presence of $O_2$.

It is noted that the optimized structures corresponding to the reaction (4), namely the production of methanol, lower the transition state energy whenever one of the atomic negative ion catalysts is used compared to the absence of a catalyst (see Figs. 2(A) through 2(J)). From the transition state calculations using dispersion-corrected density-functional theory we have found that the atomic $Ag^-$, $At^-$, $Ru^-$, $Os^-$ and $Au^-$ ions yield the transition states of 2.79, 2.88, 2.97, 2.69, and 3.01eV, respectively. Although, in our recent study we concluded that the atomic $Au^-$ ion is an excellent catalyst [1], clearly from this calculation the atomic Ag⁻, At⁻, Ru⁻, and Os⁻ ions have a higher catalytic effect on the formation of methanol from $CH_4$ without the $CO_2$ emission, with the Os⁻ ion representing the best catalyst. These results are also summarized in Table 1 from which we have selected atoms whose data are better than that for the atomic Au.

To further elucidate and understand the catalytic activity of the atomic negative ion catalysts on the oxidation of methane to the useful product such as methanol without the $CO_2$ emission, the thermodynamic properties of the possible nine reactions were analyzed using Density Functional Theory (DFT) and dispersion-corrected DFT [34] as well and the graphs of the change in the Gibbs free energy in kcal/mole versus temperature, T in Kelvin were plotted to find the optimum temperature of a specific reaction. From the Figure 3, it is clear that $\Delta G$ (= $\Delta H$ -T$\Delta S$) versus T is approximately a linear equation, particularly at higher temperatures with a slope corresponding to the change in entropy and the T-intercept indicates the optimum temperature thermodynamically favorable to the formation of the product. By controlling the temperature around 290, 300, 310, 320 and 325 K methane can be completely oxidized to methanol, rather than to carbon dioxide using the atomic Os⁻, Ag⁻, At⁻, Ru⁻ and Au⁻ ion catalysts, respectively. As seen from Table 1, these thermodynamics properties agree excellently with the transition state calculations of the oxidation of methane to methanol using the selected atomic negative ion catalysts.

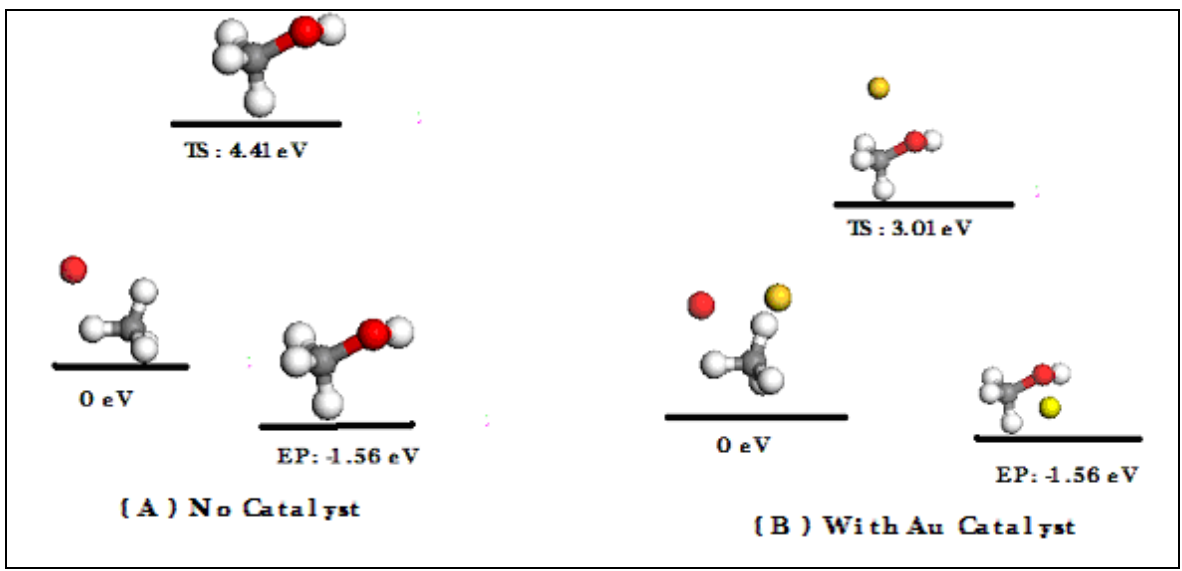

**Fig. 2 (A):** Transition state (TS), and product (EP) of the oxidation of methane to methanol in the absence of a catalyst. The red, white and grey spheres represent the O, H and C, respectively.
**Fig. 2 (B):** Transition state (TS), and product (EP) of the oxidation of methane catalyzed by the atomic Au⁻ ion to methanol. The red, gold, white and grey spheres represent the O, Au⁻ ion, H and C, respectively.

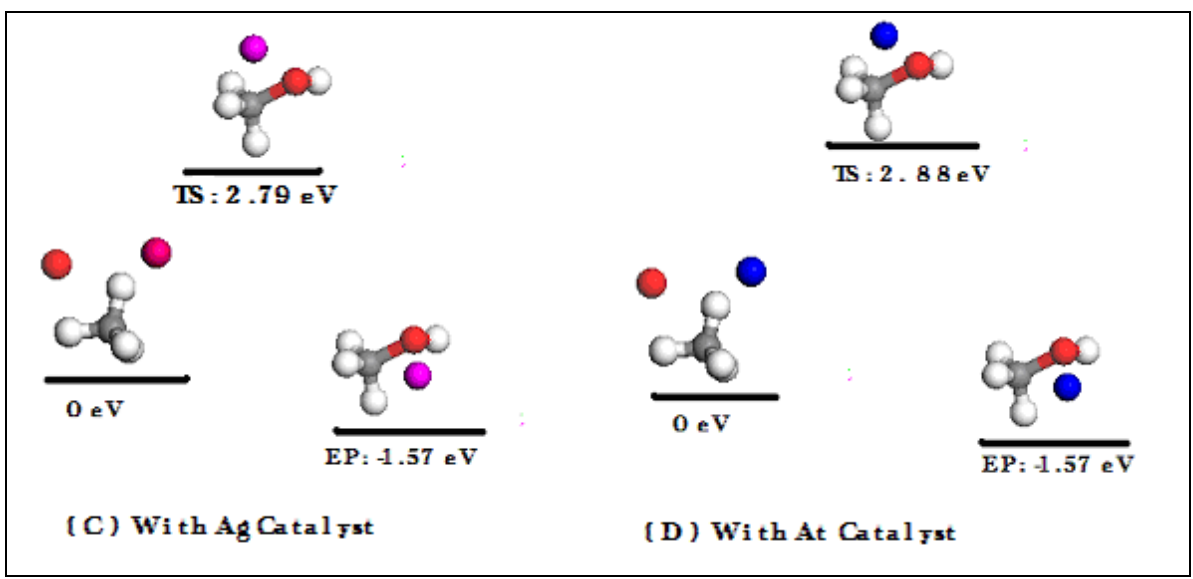

**Fig. 2 (C):** Transition state (TS), and product (EP) of the oxidation of methane catalyzed by the atomic Ag⁻ ion to methanol. The red, purple, white and grey spheres represent the O, Ag⁻ ion, H and C, respectively. **Fig. 2 (D):** Transition state (TS), and product (EP) of the oxidation of methane catalyzed by the atomic At⁻ ion to methanol. The red, blue, white and grey spheres represent the O, At⁻ ion, H and C, respectively.

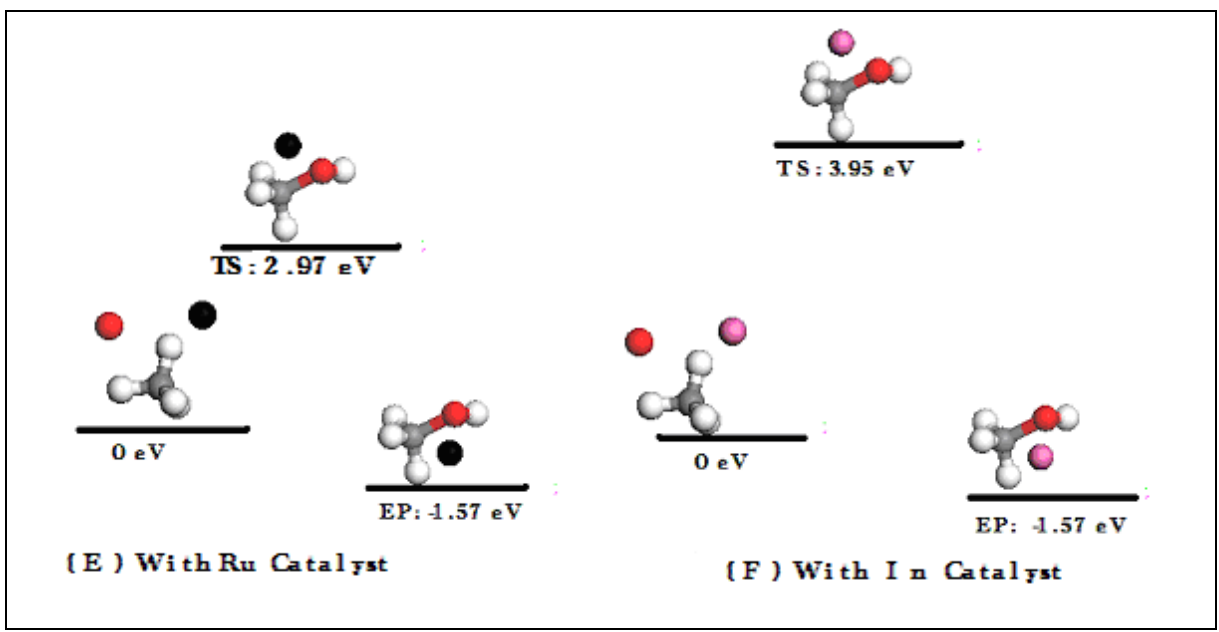

**Fig. 2 (E):** Transition state (TS), and product (EP) of the oxidation of methane catalyzed by the atomic Ru⁻ ion to methanol. The red, black, white and grey spheres represent the O, Ru⁻ ion, H and C, respectively. **Fig. 2 (F):** Transition state (TS), and product (EP) of the oxidation of methane catalyzed by the atomic In⁻ ion to methanol. The red, pink, white and grey spheres represent the O, In⁻ ion, H and C, respectively.

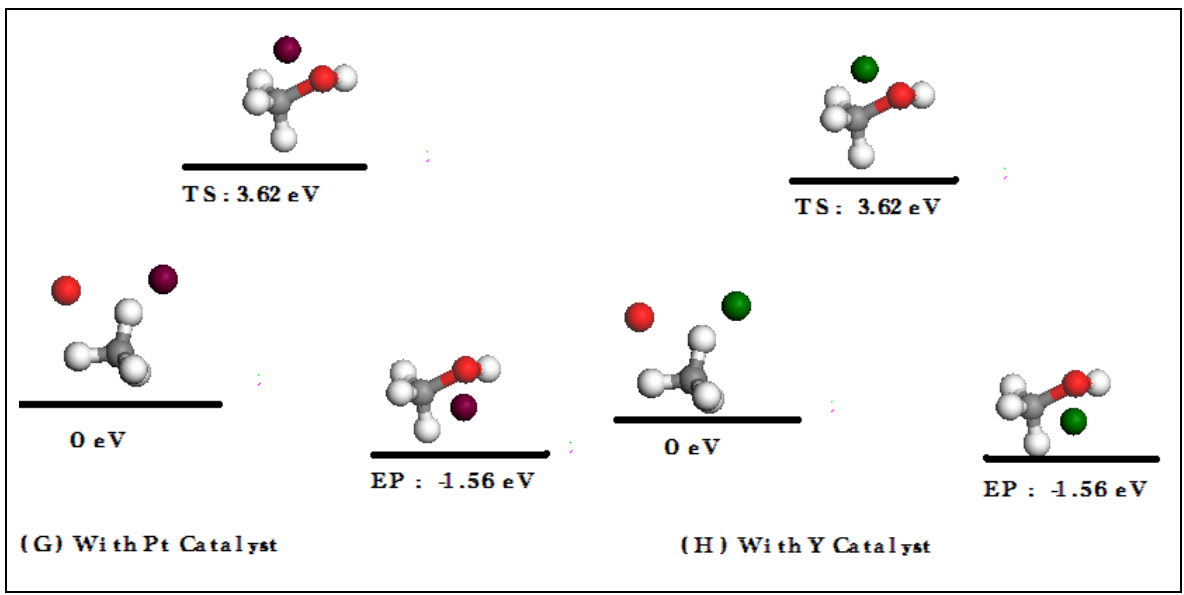

**Fig. 2 (G):** Transition state (TS), and product (EP) of the oxidation of methane catalyzed by the atomic Pt⁻ ion to methanol. The red, dark purple, white and grey spheres represent the O, Pt⁻ ion, H and C, respectively. **Fig. 2 (H):** Transition state (TS), and product (EP) of the oxidation of methane catalyzed by the atomic Y⁻ ion to methanol. The red, green, white and grey spheres represent the O, Y⁻ ion, H and C, respectively

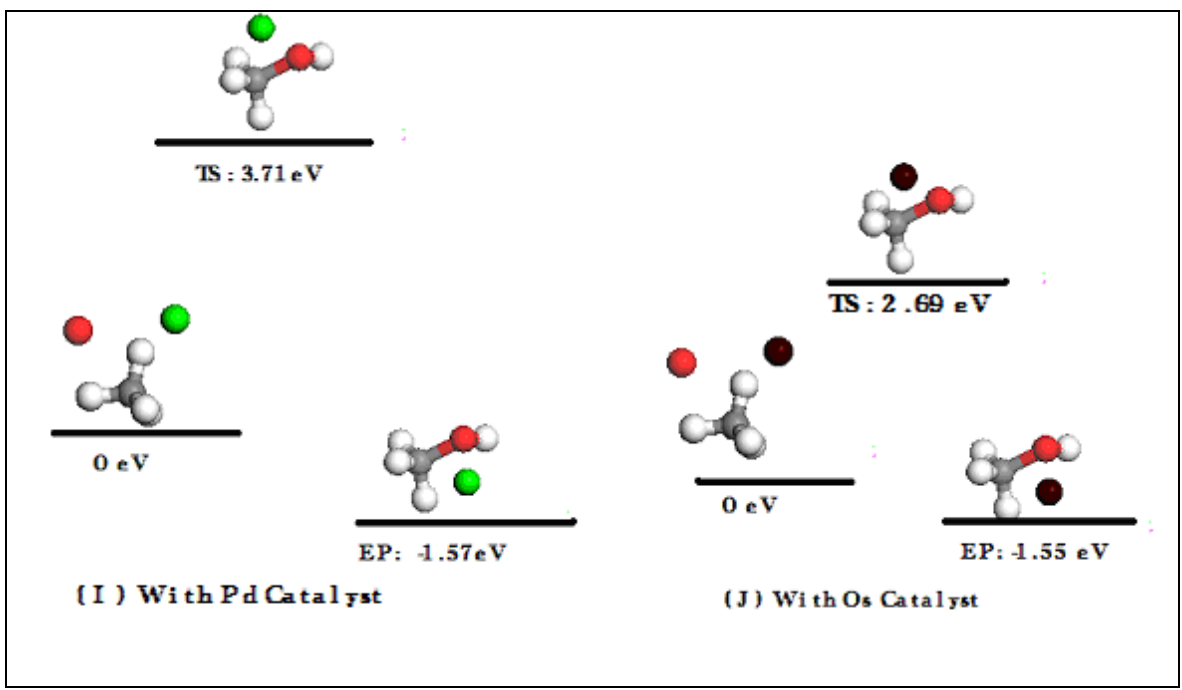

**Fig. 2 (I):** Transition state (TS), and product (EP) of the oxidation of methane catalyzed by the atomic Pd⁻ ion to methanol. The red, light green, white and grey spheres represent the O, Pd⁻

ion, H and C, respectively. **Fig. 2 (J):** Transition state (TS), and product (EP) of the oxidation of methane catalyzed by the atomic Os⁻ ion to methanol. The red, brown, white and grey spheres represent the O, Os⁻ ion, H and C, respectively.

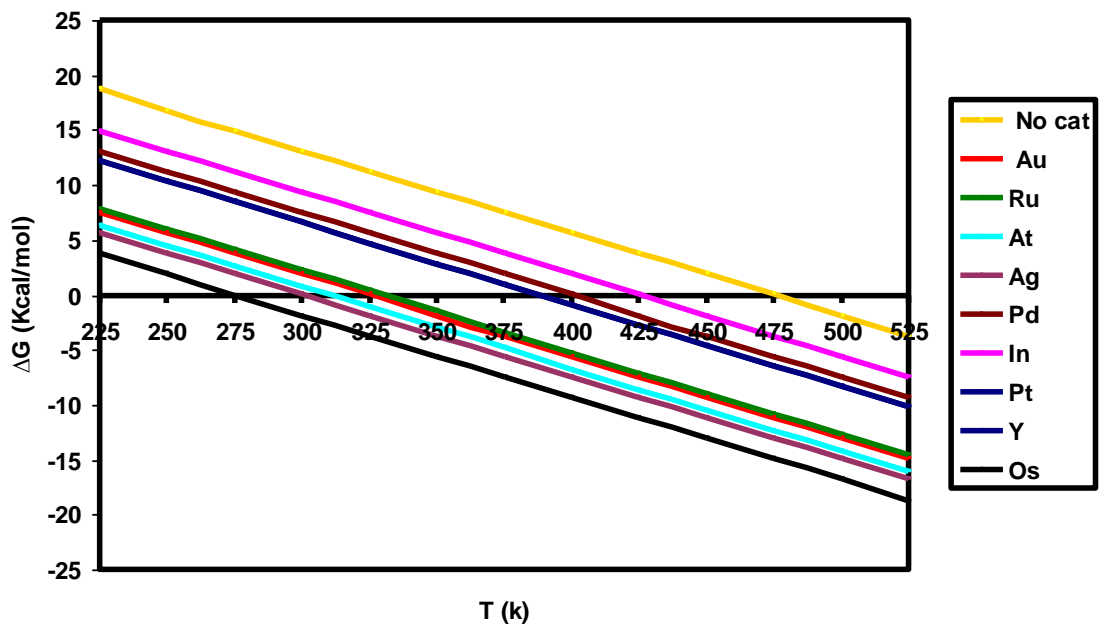

**Figure 3:** Change in the Gibbs free energy (kcal/mol) versus temperature, T (K) in the absence and presence of the Au⁻, Y⁻, Ru⁻, At⁻, In⁻, Pd⁻, Ag⁻, Pt⁻, and Os⁻ ion catalysts for the SPO of methane to methanol gas. Note that the Os⁻ ion represents the best catalyst among the nine.

### 3.1   Comment on Relativistic Effects

It is worth remarking on the significance of the relativistic effects on the calculations of the structure and the dynamics of atoms and their negative ions considered in this paper, particularly on the EA of atomic Au. Relativistic effects are known to be important in gold chemistry, Gorin and Toste [6] and Hakkinen *et al* [41] and references therein. However, accounting for relativistic effects does not necessarily guarantee reliable results, if the crucial electron-electron correlation effects and the core polarization interaction, both vital for the existence and stability of most atomic negative ions, are not adequately accounted for (see further discussions on this in [18]). Most existing theoretical methods used for calculating the binding energies (BEs) of the atomic negative ions, including the BEs of the atomic Au⁻ and Pt⁻ ions are structure-based; therefore the results obtained through these methods are often riddled with uncertainties and lack definitiveness for complex systems, such as the Au and Pt atoms. Our use of the Regge pole methodology [25] circumvents the problems associated with the structure-based theoretical methods. In the calculation of the transition barriers for $H_2O$, HDO and $D_2O$ when catalyzed by the Au⁻ ion to the corresponding peroxides, we used both the all-electron relativistic potential and the non-relativistic potential and found that relativistic effects contribution were small, less than 2.3% [18].

The above examples demonstrate that relativistic effects are not that significant in the calculation of the EAs of many atoms in general, including the atomic Au; however, electron-electron correlations and the polarization interaction are crucial. Finally, we note that when reliable binding energies of fine-structure levels were available, our results agreed very well with the weighted average value of these energies, as they should.

## 4. Summary and Conclusion

In this paper we have carried out a theoretical investigation of the catalytic activities of the atomic $Y^-$, $Ru^-$, $At^-$, $In^-$, $Pd^-$, $Ag^-$, $Os^-$ and $Pt^-$ ions for the selective partial oxidation of methane to methanol without the $CO_2$ emission. The objective was to identify effective atomic negative ion catalysts using the data for the atomic $Au^-$ ion as the benchmark. Firstly, the low energy electron scattering TCSs for the atoms of interest calculated elsewhere [30, 31, 32] were contrasted with those of the atomic Au. Then from the transition state calculations using dispersion-corrected DFT and the thermodynamics of the reactions we found that the atomic $Os^-$, $Ag^-$, $At^-$, $Ru^-$, and $Au^-$ ions lower the transition states to 2.69, 2.79, 2.88, 2.97, and 3.01 eV, respectively. The role of the atomic negative ions in catalysis is essentially to disrupt the C-H bonding in the $CH_4$ oxidation thereby eliminate the competition from the carbon dioxide formation [1]. Comparing the electron elastic scattering TCSs for the atoms Ag, At, Ru and Au, it has been found that the atoms with the lowest second R-T minima and large EA values yield the most effective negative ion catalysts (see Table 1). This demonstrates the importance of the identification and delineation of the resonance structures in the low energy electron elastic TCSs for atoms in general.

We conclude that by controlling the temperature around 290, 300, 310, 320 and 325 K methane can be completely oxidized to methanol without the emission of the $CO_2$ through the atomic $Os^-$, $Ag^-$, $At^-$, $Ru^-$ and $Au^-$ ion catalysts, respectively since in the absence of the atomic negative ion catalysts methane undergoes complete combustion at around 340 K. We also recommend the investigation of the catalytic activities of the combinations of the above negative ions for possible significant enhancement of the selective partial oxidation of methane to methanol without the $CO_2$ emission.

**Acknowledgements**

Research was supported by Army Research Office (Grant W911NF-11-1-0194) and the US DOE, Division of Chemical Sciences, Office of Basic Energy Sciences, Office of Energy Research. This research used resources of the National Energy Research Scientific Computing Center, which is supported by the Office of Science of the US DOE under Contract No. DE-AC02-05CH11231. The computing facilities of the Queen's University of Belfast, UK are also greatly appreciated.